\documentclass{icrc}

\usepackage{times}
\usepackage{graphicx} 

\begin{document}

\sloppy

\title{The Shape of EAS Lateral Distribution
and Primary Composition of the UHE Cosmic Rays}
\author[1]{R.\,I. Raikin}
\author[1]{A.\,A. Lagutin}
\affil[1]{Department of Theoretical Physics, Altai State University, Barnaul, Russia}
\author[2]{N. Inoue}
\affil[2]{Department of Physics, Saitama University, Saitama, Japan}
\author[3]{A. Misaki}
\affil[3]{Advanced Research Institute for Science and Engineering, Waseda University, Tokyo, Japan}

\correspondence{\\R.\,I.~Raikin (e-mail:~raikin@theory.dcn-asu.ru)}

\runninghead{R.\,I.~Raikin et al.: The Shape of EAS Lateral Distribution and Primary Composition \ldots}

\firstpage{1}

\pubyear{2001}

\maketitle

\begin{abstract}
Theoretical predictions for lateral distribution function of electrons in
extensive air showers based on scaling formalism are presented. Our results
are tested by comparison with AGASA experimental data taking into account
the contribution of low-energy muons and simultaneously the effect of
scintillation detectors response, according to recent simulation results.
The possibility for ultrahigh energy cosmic ray primary composition
deduction from the shape of LDF is discussed in detail.
\end{abstract}

\section{Introduction}
One of the basic characteristics of extensive air showers (EAS) of
superhigh energies, that can be measured quite accurately by large
ground-based air shower arrays is charged particle local density at various
distances from the core location. Reliable theoretical predictions on the
lateral distribution function
(LDF) of different EAS 
components in wide radial distance range are therefore
of great importance in cosmic ray research for both physical interpretation
of existing experimental data and new experiments design studies.
Unfortunately, the problem of fast, proper and at the same time adequate to
numerous specific experimental conditions calculations of LDF at large
distances from the shower core ($r\geq 1$~km) is not solved yet.

Direct Monte-Carlo technique is of course preferable but takes unreasonably
much computation time. One of widely used approaches implements an
analytical description of electromagnetic subshowers, usually based on
different well known modifications of NKG formula
\citep{dedenkomod,modnkg,plya1} obtained for distances up to several
hundreds meters from the core position. In this case formal extrapolation
to $r\sim 1$~km and farther is a source of mistakes. The weak point of
hybrid methods is the necessity of complete recalculation of prerecorded
libraries of subshowers in case of any changes in input assumptions.
Established thinning algorithms seem very promising, but still not
sufficiently effective in case of very large radial distances, because
small fraction of distant particles is tracked with very large weights
\citep{mocca2}. There are evidences [see, for example, \citet{Slovenia}],
that it is possible to improve statistical significance of thinning at
large distances by artificial limiting the weight, which particle can
obtain, but such technics are still in the stage of development.

On the other hand, present situation concerning comparisons of lateral
distributions measured by scintillation counters at Akeno and Yakutsk with
each other as well as with theoretical predictions at ultrahigh energies
($E_0\geq 10^{18}$~eV) remains mostly unsatisfactory. Though in several
recent publications [see, for example, \citet{YakUSA,NaganoInoue}]
reasonable agreement between the shapes of experimental and calculational
charged particle LDFs was achieved, the absolute values demonstrate
significant discrepancies. As a consequence shifting theoretical densities
vertically with a factor~$\sim 1.5$ is now widespread technique utilized
during comparisons of lateral distributions.
Furthermore latest theoretical studies
\citep{Kutter,Raikinsc,Lagprep2000,NaganoInoue} indicate strong influence
of the effect of scintillation detector response on the shape of LDF. It is
also worth to mention peculiarities of lateral distributions of charged
particles and muons at $E_0\geq 5\cdot 10^{18}$~eV founded by Yakutsk
group, that have not been confirmed by other experiments at this moment and
can not be explained by simulations without exotic physical assumptions
\citep{yakutskl1,YakUSA}.

In this paper we check the validity of one-parametric scaling
representation of lateral distribution of electrons established in our
earlier works by comparison with AGASA experimental data. In order to
perform adequate comparisons with experiment recent CORSIKA simulation
results about contribution of low-energy muons and energy deposition in
scintillation counters are used. We also examine here the possibility of
utilizing the experimental data about the shape of charged particle LDF for
cosmic ray primary composition deduction.

\section{Scaling formalism for LDF in cascade showers}
According to scaling formalism \citep{Raikin3,Raikin1,Raikinsc}, the
lateral distribution of electrons in both gamma- and hadron-induced cascade
showers can be represented in a form:
\begin{equation}\label{estim}
\rho_e(r;E_0,t)=\frac{N_e(E_0,t)}{R_{\rm
m.s.}^2(E_0,t)}\,F\left(\displaystyle\frac{r}{R_{\rm m.s.}(E_0,t)}\right).
\end{equation}
Here $\rho_e(r;E_0,t)$~-- electron density at radial distance $r$ from the
core in showers of primary energy $E_0$ at depth $t$ from primary particle
injection point, $N_e$~-- total number of electrons at observation level,
$R_{\rm m.s.}$~-- mean square radius (the second moment of normalized
electron LDF), which is defined~as
\begin{equation}
R_{\rm m.s.}(E_0,t)=\left[\frac{2\pi}{N_e(E_0,t)}\,
\displaystyle\int\limits_0^\infty r^2
\rho_e(r;E_0,t)\,r\,dr\right]^{1/2}.\label{rms_def}
\end{equation}
Function $F(X)$ in formula~(\ref{estim})~({\em scaling function})
is normalized LDF with respect to
variable $X=r/R_{\rm m.s.}$. It does not
depend practically on primary particle type, shower energy and age.
Besides it is not sensitive to variations of basic parameters of hadronic
interaction model.
Finally we obtain the following formula for electron density
\citep{Raikin_Brazil}:
\[
\rho_e(r)=N_e\,\frac{0.28}{R_{\rm m.s.}^2}\,\left(\frac{r}{R^{\vphantom{2}}_{\rm m.s.}}\right)^{-1.2}\left(1+
\frac{r}{R^{\vphantom{2}}_{\rm m.s.}}\right)^{-3.33}\times
\]
\begin{equation}\label{ronew}
\times\left(1+
\left(\frac{r}{10\,R^{\vphantom{2}}_{\rm m.s.}}\right)^2\right)^{-0.6}.
\end{equation}
Though the structure of function~(\ref{ronew}) is similar to modified
Linsley function, which is traditionally used for fitting experimental
LDF of charged particles measured by scintillation counters, the important
difference is that function~(\ref{ronew}) is one-parametric and
scale-invariant.

Our latest results obtained for the mean square radius of electrons can be
approximated as follows \citep{Rprep2001,contr1}:
\[
R_{{\rm m.s.}}(E_0,t)=\frac{\rho_0}{\rho(t)}\,173.0\times
\]
\begin{equation}
\times\left[0.546
+\frac{2}{\pi}~{\rm arctg}\left(\frac{t}{t_{\rm
max}+100~{\rm g/cm^2}}-1\right)\right],~{\rm m}.\label{Rms_app}
\end{equation}
Here $\rho(t)$ is air density at depth $t$, $\rho_0=1.225~\rm g/cm^3$,
$t_{\rm max}$~-- depth of maximum of average cascade curve.
Approximation~(\ref{Rms_app}) gives one-valued relation between~$t_{\rm
max}$ and normalized average lateral distribution function of electrons. It
is important that relation~(\ref{Rms_app}) is stable to variation of
hadron-air inelastic cross section and inclusive spectra of secondaries
remaining valid within the limits defined by widespread hadronic
interaction models \citep{Rprep2001,contr1}.

 \begin{figure}[t]
 \vspace*{2.0mm} 
 \centering
 \includegraphics[width=7.0cm]{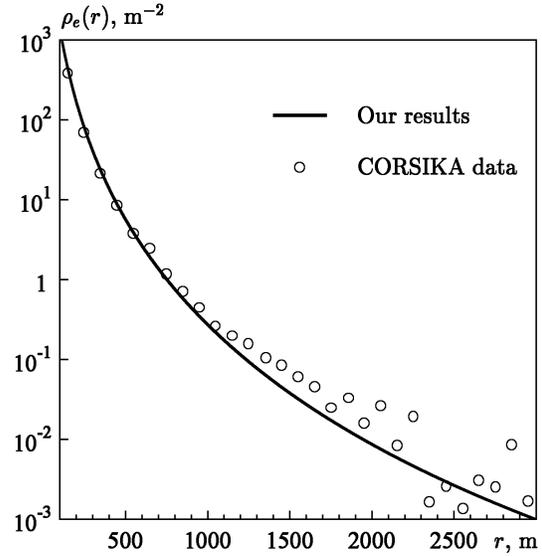} 
\caption{Average lateral distribution of electrons for $10^{18}$~eV
proton-induced extensive air showers. Function (\ref{ronew}) is shown in
comparison with CORSIKA/QGSJET data \citep{NaganoInoue}}
\label{fig1}
 \end{figure}

On Fig.~\ref{fig1} we compare our scaling LDF of electrons with
CORSIKA/QGSJET predictions \citep{NaganoInoue} for $10^{18}$~eV
proton-initiated extensive air showers. Our results are presented by
function~(\ref{ronew}) with $N_e=4.37\cdot 10^8$ and $R_{\rm m.s.}=118$~m
($t_{\rm max}=722~\rm g/cm^2$) as QGSJET predicts for proton-induced
showers of considered energy. It is seen, that function~(\ref{ronew}) is
steeper than CORSIKA data. The difference becomes essential at~$r\geq 1$~km
and increases considerably when irregularities appear in CORSIKA data due
to the lack of distant particles caused by thin sampling.

%
 \begin{figure*}[t]
 \centering
 \includegraphics[width=16.5cm]{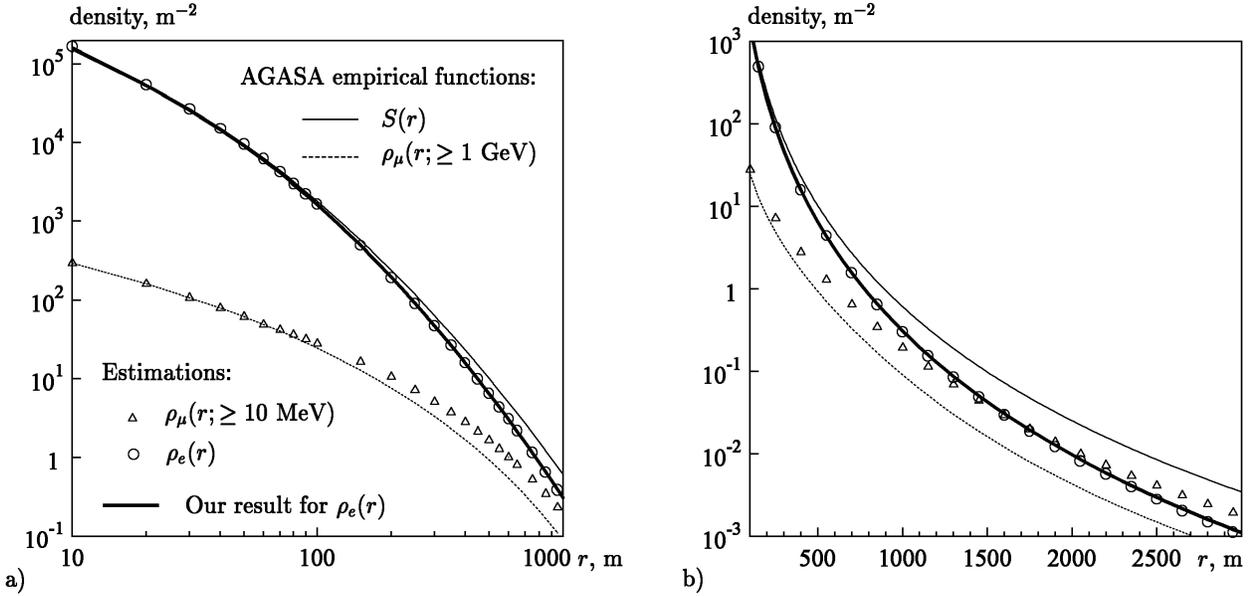}
 \caption{Comparison of our prediction for lateral distribution of electrons
 in $10^{18}$~eV proton-induced EAS with result of $\rho_e(r)$ estimation
 from AGASA data. Same data is plotted with logarithmic (a) and linear (b)
 radial distance scale}\label{fig2}
 \end{figure*}

\section{Comparison of electron LDF with AGASA data}
Adequate comparisons of theoretical results with experimental lateral
distributions of charged particles require to take into account the
contribution of low-energy muons and also detector response effect.

According to CORSIKA simulations \citep{NaganoInoue} lateral distribution
of muons changes its form sharply, when muon threshold energy decrease
from 1~GeV to 0.25~GeV. About 10-15\% additional increasing in muon
densities without disturbing the shape of LDF was pointed out for threshold
energy 10~MeV. As a result coefficient $k_\mu(r)=\rho_\mu(r;\geq 10~{\rm
MeV})/\rho_\mu(r;\geq 1~{\rm GeV})$ in vertical showers increases from
$\sim 1.8$ at 600~m to $\sim 2.7$ at 2000~m from the core making muon
lateral distribution much flatter in comparison with LDF measured
experimentally by muon detectors with high threshold energy.

In several recent papers \citep{Kutter,Raikinsc,Lagprep2000,NaganoInoue}
theoretical study of the influence of scintillation detector response on
the shape of charged particle LDF has been made.
The correction factor $k_{\rm sc}(r)=S(r)/\rho_{\rm ch}(r)$, where $S(r)$
is scintillation yield in units of minimum ionizing particle ({\em
scintillator density}) and $\rho_{\rm ch}(r)$ is charged particle density
itself, was investigated. While absolute values of $k_{\rm sc}$ calculated
using different methods and simulation codes are different from each other,
the main tendency in radial dependence is same. Function $k_{\rm sc}(r)$
increases distinctly from several tens to at least several hundreds meters
from the core position. It means that utilizing a factor 1.1 of
scintillator density to spark chamber density, which has been determined
experimentally by AGASA group within 100 m from the core \citep{ksc11}, is
not quite accurate for large distances. For our comparisons we implemented
data \citet{Kutter}, calculated by CORSIKA for 5~cm plastic scintillator,
according to which $k_{\rm sc}\sim 1.1$ at $r=50$~m and $k_{\rm sc}\sim
1.4$ at $r=600-1000$~m. For very large distances ($r\geq 1$~km) no changes
in the shape of LDF due to scintillation detector response were assumed.

Formally, the relation between two experimentally observable
characteristics $S(r)$ and $\rho_\mu(r;\geq 1~{\rm GeV})$ can be presented
as follows:
\begin{equation}
S(r)=\frac{\left[\rho_e(r)+\rho_\mu(r;\geq 1~{\rm GeV})k_\mu(r)\right]
k_{\rm sc}(r)}{K_{\rm array}}.\label{S(r)}
\end{equation}
Here $K_{\rm array}$ is coefficient, which depends on single particle
definition used in concrete experiment [in case of AGASA $K_{\rm
array}=1.1$ \citep{NaganoInoue}].

Since $k_\mu(r)$ and $k_{\rm sc}(r)$ are determined basically by low
energetic part of a cascade, it is reasonable to expect that radial
behaviour of both correction factors will not depend critically on features
of hadronic interactions at very high energies, primary energy (at least in
a limited energy range) and composition. This conclusion has been partially
confirmed by theoretical studies \cite{Kutter,Lagprep2000,NaganoInoue}.
Therefore we can use relation~(\ref{S(r)}) to estimate electron LDF from
the experimental data. On Fig.~\ref{fig2} we show the result of such
estimation of $\rho_e(r)$ for vertical $10^{18}$~eV showers obtained from
AGASA data in comparison with our scaling distribution for proton
primaries~(same as on Fig.~\ref{fig1}, except of normalizing constant).
AGASA empirical functions for $S(r)$ and $\rho_\mu(r;\geq 1~{\rm GeV})$
together with estimated $\rho_\mu(r;\geq 10~{\rm MeV})$ are also shown. We
utilized both logarithmic (a) and linear (b) radial distance scales to
emphasize that our result is in very good agreement with the one estimated
from AGASA experimental data in whole radial distance range.

\section{Cosmic ray primary composition from the shape of LDF}
The chemical composition of primary cosmic rays with $E_0\geq 10^{18}$~eV
is not measured well for today. The basic method used to deduce primary
composition from experimental data of large ground-based air shower arrays
is muon component analysis. Though muon local density at fixed distance
from the core position is rather sensitive to the type of primary particle
(including $\gamma$-rays), the observable muons with $E_{\rm
th}=(0.5-1)~{\rm GeV}\cos{\theta}$ are mostly derived from high energy
interactions. Thus the interpretation of experimental data on the basis of
comparisons with theoretical results depends on hadronic interaction model
used in calculations. This problem is still unsolved.

 \begin{figure}[t]
 \vspace*{2.0mm} 
 \centering
 \includegraphics[width=6.0cm]{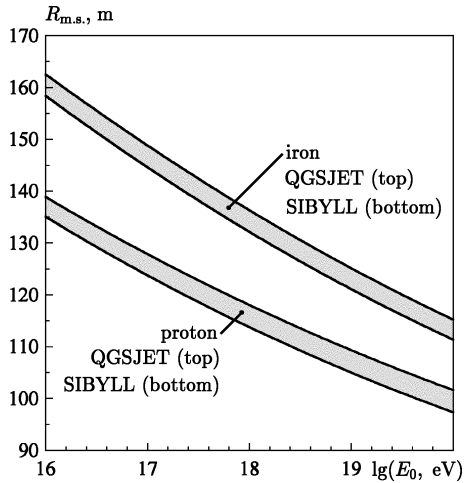} 
\caption{The energy dependence of the mean square radius of EAS electrons
for proton- and iron- initiated showers at AGASA observation depth as
predicted by approximation~(\ref{Rms_app}) with CORSIKA and SIBYLL results
for $t_{\rm max}$. See text for details}\label{fig3}
 \end{figure}

A key result, which allows us to consider the shape of lateral distribution
function measured by scintillation counters as a source of information
about primary composition of ultrahigh energy cosmic rays, is rather strong
energy (and consequently composition) dependence of the shape of electron
LDF. It is important that, as scaling property~(\ref{estim}) shows, this
dependence can be completely described by the variation of single
parameter~-- mean square radius of electrons. According to our results
\citep{Rprep2001,contr1} the narrowing of electron LDF with energy is
likely to be conserved at least up to $10^{20}$~eV, while the dependence on
basic parameters of hadronic interaction model is relatively weak. On
Fig.~\ref{fig3} we show mean square radius of electrons vs. primary energy
for proton- and iron-initiated showers at AGASA observation level ($t_{\rm
obs}=920~{\rm g/cm^2}$). The shaded areas represent uncertainties of the
result among QGSJET (top) and SIBYLL (bottom) interaction models [here we
used approximation~(\ref{Rms_app}) with these models predictions for
$t_{\rm max}$ as reported in \citet{mocca2}].

Unfortunately, the mean square radius itself can not be estimated directly
from experimental data so far as it demands high precision measurements of
lateral distribution of electron component in wide radial distance range
(from several meters to several thousands meters from the core).
Nevertheless, the change of slope of charged particle LDF around 600~m from
the shower core reflects well the changes in electron lateral distribution,
as experimental uncertainties in this region are relatively small and the
contribution of muons with $E_{\rm th}=1~\rm GeV$ in vertical showers is
about 10\% at $5\cdot 10^{18}$~eV, decreasing with energy
\citep{teshimamuons}.\balance

We should point out that theoretical studies of both $k_\mu(r)$ and $k_{\rm
sc}(r)$ are far from being complete up to now and also it is still
difficult to finally exclude some exotic processes or new particles in EAS
of ultrahigh energies. Furthermore, it is impossible to confirm or deny the
presence of small fraction of $\gamma$-induced showers from such kind of
analysis. But the main disadvantage for making some conclusions about
composition is controversial character of existing experimental data,
obtained by AGASA and Yakutsk array. While energy independence of the shape
of $S(r)$ from $\sim 10^{17.5}$~eV up to highest observed energies
reported by AGASA group repeatedly [see, for example, \citet{AGASAUSA}] is
an evidence for changing primary composition from relatively light to
relatively heavy in this energy region, Yakutsk data shows extremely sharp
steepening of LDF around $5\cdot 10^{18}$~eV \citep{yakutskl1,YakUSA},
which (refusing extraordinary explanations) leads to converse conclusion.
Nevertheless the weak sensitivity of the result to hadronic interaction
model makes such kind of analysis interesting as another one source of
information about primary composition, hopefully in nearest future with
further accumulation of experimental data.

\begin{acknowledgements}
We are grateful to K.\,Shinozaki from Saitama University for valuable
discussions. One of the authors (R.\,I.\,R.) is greatly indebted to
Japan-Russia Youth Exchange Center (JREX) for support of his research work
in Japan in form of a fellowship. He also wishes to thank Profs. A.\,Misaki
and N.\,Inoue and all members of Saitama University cosmic ray research
group for the warm hospitality.
\end{acknowledgements}





\begin{thebibliography}{99}

\bibitem[L.\,G.\,De\-den\-ko et al.(1975)]{dedenkomod}
L.\,G.\,Dedenko, N.\,M.\,Nesterova, S.\,I.\,Nikolsky et al. Proc. 14 ICRC,
M\"{u}nchen, 1975, Vol.\,8, p.\,2731.

\bibitem[A.\,Fi\-lip\-cic et al.(2000)]{Slovenia}
A.\,Filipcic, M.\,Kobal, D.\,Zavtranik. GAP-2000-028 (Auger Technical
Note), 2000.

\bibitem[A.\,V.\,Glush\-kov et al.(1997)]{yakutskl1}
A.\,V.\, Glushkov, M.\,I.\,Pravdin, I.\,Ye.\,Sleptsov. Proc. 25 ICRC,
Durban, 1997, Vol.\,6, p.\,233.

\bibitem[A.\,V.\,Glush\-kov et al.(1999)]{YakUSA}
A.\,V.\,Glushkov, M.\,I.\,Pravdin, V.\,R.\,Sleptsova et al. Proc. 26 ICRC,
Salt Lake City, 1999, Vol\,1, p.\,399.

\bibitem[A.\,I.\,Gon\-cha\-rov et al.(2000)]{Lagprep2000}
A.\,I.\,Goncharov, A.\,A.\,Lagutin, V.\,V\,Melentieva. Preprint/ASU;
2000/15, Barnaul, 2000.

\bibitem[N.\,Ha\-ya\-shi\-da et al.(1995)]{teshimamuons}
N.\,Hayashida, K.\,Honda, M.\,Honda et al. J. Phys. G: Nucl. Part. Phys.,
21, 1995, P.\,1101.

\bibitem[N.\,Ha\-ya\-shi\-da et al.(1999)]{AGASAUSA}
N.\,Hayashida, K.\,Honda, N.\,Inoue et al. Proc. 26 ICRC, Salt Lake City,
1999, Vol\,1, p.\,353.

\bibitem[A.\,M.\,Hillas(1997)]{mocca2}
A.\,M\,Hillas. Nucl. Phys. B (Proc. Suppl.), 52\,B, 1997, P.\,29.

\bibitem[T.\,Kutter(1998)]{Kutter}
T.\,Kutter. GAP-98-048 (Auger Technical Note), 1998.

\bibitem[A.\,A.\,La\-gu\-tin et al.(1979)]{modnkg}
A.\,A.\,Lagutin, A.\,V.\,Plyasheshnikov,
V.\,V.\,Uchaikin. Proc. 16 ICRC, Kyoto, 1979, Vol.\,7. p.\,18.

\bibitem[A.\,A.\,La\-gu\-tin et al.(1997)]{Raikin3}
A.\,A.\,Lagutin, A.\,Misaki and R.\,I.\,Raikin. Proc. 25 ICRC, Durban,
1997, Vol.\,6, p.\,285.

\bibitem[A.\,A.\,La\-gu\-tin et al.(1998)]{Raikin1}
A.\,A.\,Lagutin, A.\,V.\,Plyasheshnikov, V.\,V.\,Melentjeva et al. Izv.
ASU, Barnaul, 1998, p.\,33.

\bibitem[A.\,A.\,La\-gu\-tin et al.(1999)]{Raikinsc}
A.\,A.\,Lagutin, A.\,V.\,Plyasheshnikov, V.\,V.\,Melentieva et al. Nucl.
Phys. B (Proc. Suppl.), 75\,A, 1999, p.\,290.

\bibitem[A.\,A.\,La\-gu\-tin, R.\,I.\,Rai\-kin(2001)]{Raikin_Brazil}
A.\,A.\,Lagutin, R.\,I.\,Raikin. Nucl. Phys. B
(Proc. Suppl.), 2001, V.97\,B, p.\,274.

\bibitem[A.\,A.\,La\-gu\-tin et al.(2001)]{Rprep2001}
A.\,A.\,Lagutin, R.\,I.\,Raikin, N\,Inoue, A.\,Misaki. Preprint/ASU;
2001/1, Barnaul, 2001.

\bibitem[M.\,Na\-ga\-no et al.(2000)]{NaganoInoue}
M.\,Nagano, D\,Heck, K\,Shinozaki et al. Astropart. Phys., Vol.\,13, 2000,
p.\,277.

\bibitem[A.\,V.\,Plya\-shesh\-ni\-kov et al.(1988)]{plya1}
A.\,V.\,Plyasheshnikov, A.\,K.\,Konopelko, K.\,V.\,Vorobjev. Preprint/\-FIAN;
No\,92, Moscow, 1988.

\bibitem[R.\,I.\,Rai\-kin et al.(this proccedings)]{contr1}
R.\,I.\,Raikin, A.\,A.\,Lagutin, N\,Inoue, A.\,Misaki. This proceedings.

\bibitem[M.\,Te\-shi\-ma et al.(1986)]{ksc11}
M.\,Teshima, Y.\,Matsubara, T\,Hara et al. J. Phys. G: Nucl. Phys., 12, 1986,
P.\,1097.

\end{thebibliography}
\end{document}